\documentclass[english,floatfix,showkeys,superscriptaddress,nofootinbib,aps,prd]{revtex4}
\usepackage[latin1]{inputenc}
\usepackage[T1]{fontenc}
\usepackage{lmodern}
\usepackage{amsmath}
\usepackage{amssymb}
\usepackage{graphicx}
\usepackage{esint}
\usepackage{longtable}
\usepackage{dcolumn}
\usepackage{babel} 
\usepackage{csquotes} 
\usepackage{color} 

\begin{document}

\title{Black holes and wormholes in Deser--Woodard gravity}

\author{Juliano C. S. Neves} 
\email{juliano.c.s.neves@gmail.com}
\affiliation{Instituto de Ciência e Tecnologia, Universidade Federal de Alfenas, \\ Rodovia José Aurélio Vilela,
11999, CEP 37715-400 Poços de Caldas, MG, Brazil}

\begin{abstract}
The Deser--Woodard gravity is a modified theory of gravity in which nonlocality plays a central role.
In this context, nonlocality is a consequence of the inverse of the d'Alembertian operator $\square^{-1}$ in the 
effective action. Here, exact black hole and wormhole solutions are built in the revised 
Deser--Woodard gravity following
a recent approach, where a special tetrad frame simplifies the complicated field equations of the theory. 
Using the Schwarzschild metric and the Reissner--Nordström metric 
as initial seed solutions, the developed algorithm generates new traversable wormholes, singular black holes 
and a regular black hole as solutions of the vacuum field equations of the modified theory.
Also, the auxiliary fields, which are responsible for the nonlocality, are computed. However, 
even for a regular black hole solution, 
in which spacetime does not contain a curvature singularity, the corresponding auxiliary fields diverge at the event horizon. Regarding observational results, the shadow angular radius is computed for the new solutions. In particular,
the deviation of the Schwarzschild black hole in the Deser--Woodard gravity 
 casts a larger shadow than the corresponding black hole in general relativity. 
\end{abstract}


\maketitle

\section{Introduction}
\label{Intro}
Nonlocal theories of gravity are hypothetical frameworks that seek to overcome the limitations of general relativity \cite{Deser:2007jk,Deser:2019lmm,Nojiri:2007uq}.
Nonlocality is an inherent feature of quantum mechanics\footnote{At least for the 
standard quantum mechanics interpretation \cite{Jammer}. See Ref. \cite{Capozziello:2021krv}
 for a review on nonlocality in gravitation.}  and, in some sense, is synonymous with
 action at a distance. Arguably, by imposing nonlocality in the gravitational sector would avoid 
instabilities of ultraviolet extensions of general relativity towards a complete quantum theory of gravity \cite{Modesto:2017sdr,Knorr:2018kog,Buoninfante:2018mre}.  

The  first Deser--Woodard nonlocal theory \cite{Deser:2007jk} aims to generate accelerating cosmic 
expansion without a cosmological constant. Using a field $X= \square ^{-1}R$ in an effective action, where
$\square^{-1}$ is the inverse of the d'Alembertian operator and $R$ is the Ricci scalar, the authors
state that this field would carry current effects of primordial infrared gravitons, thereby explaining the current
cosmic acceleration \cite{Deser:2019lmm}.
Later, Belgacem \textit{et al.} \cite{Belgacem:2018wtb}
 showed a fatal flaw of the first Deser--Woodard model: the effects of nonlocality
would be detectable in solar system experiments, something in disagreement  with the data.
In order to overcome this situation, the second or the revised Deser--Woodard  theory \cite{Deser:2019lmm} 
adopts an extra field $Y= \square^{-1}g^{\mu\nu}\partial X_{\mu}\partial X_{\nu}$ 
that maintains the virtue of the first model and avoids the issues of the solar system experiments.

In the context of the Deser--Woodard second theory, studies in cosmology \cite{Ding:2019rlp,Jackson:2023faq,Chen:2019wlu,Jackson:2021mgw} and in gravitation have been 
made \cite{Chen:2021pxd,DAgostino:2025sta,DAgostino:2025wgl}.\footnote{In Ref. \cite{Chen:2021pxd}, quasinormal modes of the Schwarzschild spacetime (not a deviation) are calculated.}
In gravitation, analytic solutions are difficult due to the nonlinearity of the field equations.
Recently, D'Agostino and De Falco \cite{DAgostino:2025sta} 
overcome this difficulty using a special tetrad frame in order
to build analytically wormhole solutions. The path is the adoption of the tetrad frame used by Morris and Thorne 
in the famous work about wormholes \cite{MT}. The special tetrad turns the field equations into a simpler form.
Following this approach, D'Agostino and De Falco \cite{DAgostino:2025wgl}, in a second paper,
 also attempted to obtain analytic black hole solutions.
However, the spacetime metrics calculated by the authors are an approximation in the perturbation parameter.

In this article, following  
D'Agostino and De Falco \cite{DAgostino:2025sta,DAgostino:2025wgl}, fully analytic black hole and wormhole solutions are calculated in the revised Deser--Woodard theory.
These metrics are solutions of the vacuum field equation and 
are deviations or deformations of either the Schwarzschild or the Reissner--Nordström
spacetime. The solutions are divided into three classes: black holes, wormholes and naked singularities.
This article will focus on black holes and wormholes. Interestingly, from the approach adopted,
deviations from the Schwarzschild metric and the Reissner--Nordström metric produce singular 
black holes and wormholes, but
a regular black hole is generated just from the extremal Reissner--Nordström metric 
(in the extremal case, electric charge equals mass). 

Regular black holes \cite{Bardeen,Neves:2021dqx,Neves:2019ywx,Neves:2015zia,Rodrigues:2015ayd,Neves:2015vga,Pedrotti:2024znu,Maeda:2021jdc,Simpson:2021dyo,Maluf:2018ksj,Maluf:2018lyu,Ghosh:2014hea,Lobo:2020ffi,Toshmatov:2014nya,Rodrigues:2018bdc,Lemos:2011dq,Neves:2014aba,Bambi:2013ufa,Molina:2010yu,Hayward:2005gi,Dymnikova:1992ux,Bronnikov:2000vy,Calza:2024fzo,Calza:2024xdh,Calza:2025mwn} and wormholes \cite{Visser:1989kh,Lemos:2003jb,Simpson:2018tsi,Lobo:2009ip,Harko:2013yb,Kanti:2011jz,Agnese:1995kd,Moraes:2017mir,Ovgun:2018xys,Nicolini:2009gw,Kanti:2011yv,Santos:2023zrj,Frizo:2022jyz,Neves:2024zwi,Molina:2012ay,Casadio:2001jg,Bronnikov:2003gx,Lobo:2007qi,Barcelo:2000ta,Klen:2020kdb,Klen:2025ush,DeFalco:2021klh,Lu:2024prz}
 have been studied in the general relativity context and beyond.
In general, both classes of spacetime metrics violate energy conditions. 
For example, the weak energy violation removes the singularity in regular black holes \cite{Neves:2014aba}. 
On the other hand, exotic matter should support the wormhole throat \cite{Visser:1989kh,Barcelo:2000ta}.
When both the regular black hole and the wormhole solution obtained here are studied in the
general relativity context with an effective energy-momentum tensor, we see the weak violation
and some sort of exotic matter that supports these metrics. 

In the last section of the paper, the shadow angular radius of the mentioned solutions is 
calculated. Since the historical EHT collaboration paper on the M87* shadow \cite{EventHorizonTelescope:2019dse}, 
black hole and wormhole shadows have become a useful tool to test metric candidates even in 
modified theories of gravity \cite{Bambi:2019tjh,Vagnozzi:2019apd,Allahyari:2019jqz,Khodadi:2020jij,Kumar:2020owy,Neves:2020doc,Neves:2019lio,Vagnozzi:2022moj,Maluf:2020kgf}.\footnote{In
recent works \cite{EventHorizonTelescope:2021dqv,Afrin:2023uzo,Afrin:2022ztr,EventHorizonTelescope:2022xqj}, regular black hole and wormhole metrics have been adopted to fit (or not) the recent
 data from the EHT collaboration. Interestingly, in spite of the Kerr hypothesis being adopted and favored, 
 not even the wormhole spacetime was ruled out for the Sgr A* data \cite{EventHorizonTelescope:2022xqj,Vagnozzi:2022moj}.}
 In particular, as the Deser--Woodard gravity is a modified theory candidate,
I show that the deviation from the Schwarzschild metric obtained casts a larger shadow than the standard
metric in general relativity.

The structure of this article is as follows: in Sec. \ref{Sec-II} the field equations of 
the Deser--Woodard gravity are presented, 
and the auxiliary fields are defined. In Sec. \ref{Sec-III} an approach to solve the field equations and
classify spacetime solutions is proposed. 
Sec. \ref{Sec-IV} and Sec. \ref{Sec-V} deal with deviations from the Schwarzschild and 
Reissner--Nordström spacetimes, respectively, in the context studied. In Sec. \ref{Sec-VI} one focuses on two solutions of 
the Deser--Woodard gravity: a Morris--Thorne wormhole and a regular black hole. For these
two solutions besides a singular black hole, 
the shadow angular radius is calculated and compared with the Schwarzschild case in 
Sec. \ref{Sec-VII}.  The final remarks are in Sec. \ref{Sec-VIII}. In this article, one adopts Planck units, i.e., $G=c=k_B=h=1$, where $G$ is the gravitational constant, $c$ is the speed of light in vacuum, 
$k_B$ is the Boltzmann constant, and $\hslash$ is the reduced Planck constant.

\section{Deser--Woodard gravity}
\label{Sec-II}
The action in the revised Deser--Woodard theory \cite{Deser:2019lmm} is given by
\begin{equation}
S= \frac{1}{16 \pi} \int d^4x \sqrt{-g}R[1+f(Y)], 
\label{action}
\end{equation}
where $g$ is the metric determinant, $R$ is the Ricci scalar or scalar curvature, and the function $f(Y)$ encodes the 
nonlocal effects of the theory. Such a function is also called the distortion function.
According to Deser and Woodard \cite{Deser:2019lmm}, the localized form of the action (\ref{action}) is obtained from 
auxiliary fields $(X,Y,U,V)$,
whose field equation are
\begin{subequations}
\begin{align}
\square X & =  R, \label{X}\\
\square Y  & =  g^{\mu\nu}\partial_{\mu} X \partial_{\nu} X, \label{Y}\\
\square  U & =  -2 \nabla_{\mu }(V \nabla^{\mu}X), \label{U}\\
\square V & =   R \frac{df}{dY}, \label{V}
\end{align}
\end{subequations}
where $\nabla_{\mu}$ is the covariant derivative, and $\square \equiv \nabla_{\mu} \nabla^{\mu}$ 
is the d'Alembertian, which for a given function $u$ is written as 
\begin{equation}
\square u \equiv \frac{1}{\sqrt{-g}}\partial_{\mu}\left[\sqrt{-g} \ \partial^{\mu}u \right].
\end{equation}

Varying the action (\ref{action}) with respect to the metric $g_{\mu\nu}$ give us the vacuum field equations of the 
Deser--Woodard theory of gravity, namely
\begin{equation}
 \left(G_{\mu\nu} + g_{\mu\nu} \square -\nabla_{\mu}\nabla_{\nu} \right) [1+U+f(Y)] + \mathcal{K}_{(\mu\nu)}  -\frac{1}{2}g_{\mu\nu}g^{\alpha \beta} \mathcal{K}_{\alpha \beta}=0,
 \label{field_equations}
\end{equation}
in which the tensor $\mathcal{K}_{\mu\nu}$ depends on the auxiliary fields, i.e.,
\begin{equation}
\mathcal{K}_{\mu\nu}= \partial_{\mu}X\partial_{\nu}U+\partial_{\mu}Y\partial_{\nu}V+V\partial_{\mu}X\partial_{\nu}X,
\label{k}
\end{equation}
with $\mathcal{K}_{(\mu\nu)}=\frac{1}{2}(\mathcal{K}_{\mu\nu}+\mathcal{K}_{\nu\mu})$.

Here I am interested in static solutions of the field equations (\ref{field_equations}) that describe, for example,
static or nonrotating black holes and wormholes. The general metric \textit{Ansatz} for static solutions 
in the $(t,r,\theta,\phi)$ coordinates is given by
\begin{equation}
ds^2 = - A(r)dt^2 + \frac{dr^2}{B(r)} +r^2 (d\theta^2 + \sin^2 \theta d\phi^2).
\label{ansatz}
\end{equation}
In order to solve the field equations (\ref{field_equations}) for this metric \textit{Ansatz}, I follow
D'Agostino and De Falco \cite{DAgostino:2025sta,DAgostino:2025wgl}
 by adopting a suitable tetrad associated with static observers. In this 
tetrad, the metric tensor is the Minkowski metric, $\eta_{\mu\nu}=\text{diag}(-1,1,1,1)$. That is, adopting the tetrad
\begin{equation}
\text{e}_{\hat{t}}=\frac{\text{e}_t}{\sqrt{A(r)}}, \ \text{e}_{\hat{r}}=\frac{\text{e}_r}{\sqrt{B(r)}}, \ \text{e}_{\hat{\theta}}=\frac{\text{e}_\theta}{r}, \ \text{e}_{\hat{\phi}}=\frac{\text{e}_\phi}{r \sin \theta},
\label{tetrad}
\end{equation}
one has $g_{\hat{\mu}\hat{\nu}}=e_{\hat{\mu}}^{\alpha}e_{\hat{\nu}}^{\beta}g_{\alpha\beta} = \eta _{\mu\nu}$.
As we will see, adopting this tetrad makes the field equations easier. Moreover, tensor components could be
simpler in this tetrad than in the coordinate basis. Other tensors in the field equations written in this tetrad become
\begin{subequations}
\begin{align}
G_{\hat{\mu}\hat{\nu}} & = e_{\hat{\mu}}^{\alpha}e_{\hat{\nu}}^{\beta}G_{\alpha \beta}=R_{\hat{\mu}\hat{\nu}}-\frac{1}{2}\eta_{\mu\nu}R, \\
\nabla_{\hat{\mu}}\nabla_{\hat{\nu}} & = e_{\hat{\mu}}^{\alpha}e_{\hat{\nu}}^{\beta}\nabla_{\alpha}\nabla_{\beta},\\
\mathcal{K}_{\hat{\mu}\hat{\nu}}& = e_{\hat{\mu}}^{\alpha}e_{\hat{\nu}}^{\beta}\mathcal{K}_{\alpha\beta},
\end{align}
\end{subequations}
where the transformation matrix and its inverse are 
\begin{equation}
e_{\hat{\mu}}^{\alpha} =  \text{diag} \left(\frac{1}{\sqrt{A(r)}},\sqrt{B(r)},\frac{1}{r},\frac{1}{r \sin \theta} \right),
\end{equation}
and
\begin{equation}
e_{\alpha}^{\hat{\mu}} = \text{diag} \left(\sqrt{A(r)},\frac{1}{\sqrt{B(r)}}, r, r \sin \theta \right).
\end{equation}

In the tetrad (\ref{tetrad}), the field equations are rewritten as
\begin{equation}
 \left(G_{\hat{\mu}\hat{\nu}} + \eta_{\mu\nu} \square -\nabla_{\hat{\mu}}\nabla_{\hat{\nu}} \right) W + \mathcal{K}_{(\hat{\mu}\hat{\nu})} -\frac{1}{2}\eta_{\mu\nu}\eta^{\alpha \beta} \mathcal{K}_{\hat{\alpha} \hat{\beta}}=0,
\end{equation}
where the function $W$ is defined as
\begin{equation}
W(r) = 1 + U(r) + f(Y(r)).
\label{f}
\end{equation}
From the metric \textit{Ansatz} (\ref{ansatz}), the auxiliary fields are radial ones, and the field equations
present only three independent components, that is, 
\begin{subequations}
\begin{align}
 \left(G_{\hat{t}\hat{t}} - \square -\nabla_{\hat{t}}\nabla_{\hat{t}} \right) W + \frac{1}{2}\mathcal{K}_{\hat{r} \hat{r}}& = 0, \label{Gtt}\\
 \left(G_{\hat{r}\hat{r}} + \square -\nabla_{\hat{r}}\nabla_{\hat{r}} \right) W + \frac{1}{2}\mathcal{K}_{\hat{r} \hat{r}}& = 0, \label{Grr}\\
 \left(G_{\hat{\theta}\hat{\theta}} + \square -\nabla_{\hat{\theta}}\nabla_{\hat{\theta}} \right) W - \frac{1}{2}\mathcal{K}_{\hat{r} \hat{r}} & = 0. \label{Gtheta}
\end{align}
\end{subequations}
Following D'Agostino and De Falco \cite{DAgostino:2025sta,DAgostino:2025wgl}, 
we should combine the three independent equations above to get rid of 
$\mathcal{K}_{\hat{r} \hat{r}}$. That is, combing (\ref{Gtt}) with (\ref{Gtheta}) and (\ref{Grr})
 with (\ref{Gtheta}), one has
\begin{subequations}
\begin{align}
&\left(G_{\hat{t}\hat{t}} + G_{\hat{\theta}\hat{\theta}} \right) W = \left(\nabla_{\hat{t}}\nabla_{\hat{t}} + \nabla_{\hat{\theta}}\nabla_{\hat{\theta}}  \right) W, \label{G1}\\
& \left(G_{\hat{r}\hat{r}} + G_{\hat{\theta}\hat{\theta}} \right) W + 2\square W = \left(\nabla_{\hat{r}}\nabla_{\hat{r}} + \nabla_{\hat{\theta}}\nabla_{\hat{\theta}}  \right) W.\label{G2}
\end{align}
\end{subequations}
One alternative to solve the system of equations (\ref{G1})-(\ref{G2}) is choosing a function $A(r)$ and then one obtains 
$W(r)$ and $B(r)$. This is the alternative of  D'Agostino and De Falco \cite{DAgostino:2025sta,DAgostino:2025wgl}.

However, interestingly, choosing the metric 
\textit{Ansatz} (\ref{ansatz}) and the constraint
\begin{equation}
W(r) = \frac{1}{A(r)},
\label{W}
\end{equation}
the two equations (\ref{G1}) and (\ref{G2}) become equal, i.e.,
\begin{equation}
A''-\frac{3 A'^2}{2 A}+ \left(\frac{B'}{2B}+\frac{3}{r} \right)A' - \left(\frac{B'}{B}-\frac{2}{rB} +\frac{2}{r} \right)\frac{A}{r} =0,
\label{final_eq}
\end{equation}
where $'$ means derivative with respect to the radial coordinate $r$.
Then, from a suitable function $A(r)$, the problem is just
calculating $B(r)$, which will be a deviation from standard solutions like the 
Schwarzschild and Reissner--Nordström solutions in this article. 

It is worth pointing out a caveat of this approach. Eqs. (\ref{G1}) and (\ref{G2}) do not necessarily
come from vacuum field equations. If  Eqs. (\ref{Gtt})-(\ref{Gtheta}) 
had  a matter field term with energy-momentum tensor given
by $T_{\hat{\mu}\hat{\nu}}=\text{diag}(T_{\hat{t}\hat{t}},T_{\hat{r}\hat{r}},T_{\hat{\theta}\hat{\theta}},T_{\hat{\phi}\hat{\phi}})$, components like $T_{\hat{t}\hat{t}}= - T_{\hat{\theta}\hat{\theta}}$ and
 $T_{\hat{r}\hat{r}}= - T_{\hat{\theta}\hat{\theta}}$ would produce the same equations (\ref{G1}) and (\ref{G2}).
 That is the reason why it is important, after obtaining the metric terms, to check the value 
 of $\mathcal{K}_{\hat{r}\hat{r}}$ or $\mathcal{K}_{rr}$ from the auxiliary fields and Eq. (\ref{k}) in order to
 make sure that Eqs. (\ref{G1}) and (\ref{G2}) come from the vacuum field equations. 

With the metric functions known, the auxiliary fields (\ref{X})-(\ref{V}) are calculated. 
From the metric \textit{Ansatz} (\ref{ansatz}), the d'Alembertian operator for a generic function $u(r)$ reads
\begin{equation}
\square u =  Bu'' + \frac{1}{2} \left (\frac{B A'}{A} + B' + \frac{4B}{r} \right)u'.
\end{equation}
After calculating the scalar curvature $R$, the fields $X$ and $Y$ are straightforwardly obtained.
Rewriting Eq. (\ref{U}), one has
\begin{equation}
\nabla_{\mu} \left(\nabla^{\mu} U + 2V \nabla^{\mu} X\right) =0  \ \   \Rightarrow \ \  U'=-2 VX'.
\label{U_line}
\end{equation}
Also, from Eq. (\ref{W}), the distortion function is $f = W-U-1$, which leads to
\begin{equation}
\frac{df}{dY} = \frac{f'}{Y'} = \frac{W' - U'}{Y'}.
\label{dfdy}
\end{equation} 
Therefore, with the aid of (\ref{U_line}) and (\ref{dfdy}), the auxiliary field $V$ is written as
\begin{equation}
\square V = \left(\frac{W'- 2VX'}{Y'} \right) R,
\end{equation}
which provides $V$ and, consequently, $U$ by using Eq. (\ref{U_line}). 
As a consequence of this procedure, the distortion function $f(Y)$ is determined.

\section{The algorithm}
\label{Sec-III}
As mentioned, the main idea is to solve (\ref{final_eq}) and try to build black hole and wormhole solutions
in the Deser--Woodard gravity. As the desired solutions are deviations or deformations from
standard solutions, I am looking for 
\begin{equation}
B(r) = \delta(r) A(r).
\label{B_delta}
\end{equation}
By using the above deviation in Eq.(\ref{final_eq}) one has
\begin{equation}
\delta (r) = \frac{A \left(4 r^2 + C A \right )}{r^2 \left(r A' - 2 A \right)^2}.
\label{delta}
\end{equation}
It is worth noting the importance of the integration constant $C$ to obtain different spacetime geometries.
Also, zeros of the numerator and denominator of (\ref{delta}) are important in the \textit{taxonomy} proposed here. 

In order to quantify the results, let us assume a generic form for the metric 
function $A(r)$ that satisfies three conditions:
\begin{itemize}
\item spacetime is asymptotically flat, that is to say, $\lim_{r \rightarrow \infty} A(r)=\lim_{r \rightarrow \infty} B(r)=1$.
\item $A(r_+) =0$, thus $r=r_+$ is the event horizon radius.
\item $A(r)$ is positive and monotonically crescent for $r>r_+$.
\end{itemize} 
Therefore, a generic form for the metric function $A(r)$ in agreement with the three items listed above 
is given by
\begin{equation}
A(r) = 1 + \sum_{n=1}^{\infty}\frac{a_n}{r^n}.
\end{equation}
For the Schwarzschild metric $a_1=-2M$, where $M$ is the mass term, and for the 
Reissner--Nordström spacetime $a_1=-2M$ and $a_2=Q^2$, where $Q$ is the electric charge.
From the event horizon assumption, one has a useful relation
\begin{equation}
 \sum_{n=1}^{\infty}\frac{a_n}{r_+^n} = -1.
 \label{minus_one}
\end{equation}
Considering $r=r_0$ root of the denominator of (\ref{delta}), thus it is straightforward to show that
\begin{equation}
r_0 = \frac{2 A(r_0)}{A'(r_0)} = 2\left( \frac{ \sum \frac{a_n}{r_+^n} -  \sum \frac{a_n}{r_0^n}}{ \sum \frac{n a_n}{r_0^{n+1}}}\right),
\label{r0}
\end{equation}
where the relation (\ref{minus_one}) was used. Rearranging terms of  (\ref{r0}) leads to 
\begin{equation}
\sum_{n=1}^{\infty}\frac{(n+2) a_n}{r_0^n} =  \sum_{n=1}^{\infty}\frac{2 a_n}{r_+^n} \ \ \ \therefore \ \ \ r_0 > r_+.
\end{equation} 
For $r=r_0$ the function $B(r)$ diverges and, consequently, the spacetime geometry does. By calculating the 
Kretschmann scalar, which depends on the Riemann tensor, 
\begin{equation}
R_{\mu\nu\alpha\beta}R^{\mu\nu\alpha\beta} =  \left(\frac{ A'^2}{4}+\frac{2 A^2}{r^2}\right) \delta'^2 + \left(A' A'' + \frac{4 A A'}{r^2}\right) \delta \delta'  + \left(A''^2+\frac{4 A'^2}{r^2}+\frac{4 A^2}{r^4}\right) \delta^2  -\frac{8 A \delta }{r^4}+\frac{4}{r^4},
\end{equation}
the curvature singularity is present both at $r=0$ and at $r=r_0$. In order to build black hole and wormhole 
solutions in this context,
the numerator of (\ref{delta}) must eliminate the singularity at $r_0$. 

\subsection{$C=C_0$}
The numerator of Eq. (\ref{delta}) provides the value of $C$ that avoids the singular point at $r=r_0$. That is,
by using Eq. (\ref{r0}), the numerator leads to 
\begin{equation}
C_0 = - \frac{16A(r_0)}{A'(r_0)^2},
\label{C0}
\end{equation}
and then spacetime is regular for $r > 0$. However, when $r<r_+$ and $A<0$, assuming that $r_0>r_+$ and
$A(r_0)>0$, the numerator of (\ref{delta}) reads
\begin{equation}
4r^2 + \frac{16A(r_0)A(r)}{A'(r_0)^2}>0.
\end{equation}
Thus, $B>0$ and the metric is not Lorentzian anymore, that is, the metric signature is not
$(-,+,+,+)$ for $r < r_+$. Indeed, the metric becomes Euclidean for this interval. 
Therefore, when $C=C_0$ with $A<0$ and $B>0$ for $r<r_+$, 
the validity of the $(t,r,\theta,\phi)$ coordinates is ensured just in the interval
$r>r_+$. 

Due to the fact that $B(r)$ has a double root at $r=r_+$, the metric extension for $r<r_+$ is 
available in this case with the aid of the quasiglobal coordinate $u$, which is 
defined by the condition $g_{tt}g_{uu}=-1$. 
The quasiglobal coordinate provides to general spherical solutions
the good and the bad properties of the radial coordinate from the Schwarzschild metric \cite{Bronnikov:2003gx,Molina:2010yu}. 
Consequently, the \textit{Ansatz} (\ref{ansatz}) is rewritten as
\begin{equation}
ds^2 = - \mathcal{A}(u)dt^2 + \frac{du^2}{\mathcal{A}(u)} +r^2 (d\theta^2 + \sin^2 \theta d\phi^2),
\label{quasiglobal}
\end{equation}  
with the following definitions:
\begin{equation}
\mathcal{A}(u)=A(r), \ \ \  r(u)=r, \ \ \  \mathcal{A}(u) \left(\frac{dr}{du} \right)^2= B(r).
\label{QG_properties}
\end{equation}
From this coordinate transformation, the metric (\ref{quasiglobal}) is Lorentzian even inside the event horizon, 
and a singular black hole is possible in this case where $C=C_0$. 

But if $A>0$ and $B>0$ for $r<r_+$, 
there exists a real root for $B(r)$ at $r=r_-<r_+$. From the numerator of
(\ref{delta}) and the value of $C$ in Eq. (\ref{C0}), we can see that $r_-$ is positive
\begin{equation}
r_{-}=\frac{2\sqrt{A(r_0)A(r_-)}}{A'(r_0)}.
\end{equation} 
On the other hand, for $r<r_-$, $B<0$ and $A>0$, thus the metric is not Lorentzian.
In this sense, the radial coordinate range is $r_- \leq r < \infty $, and the maximal extension of the metric (using
the standard procedure with the tortoise coordinate) is
viable for this interval. The spacetime geometry is regular. And as we have 
an event horizon at $r_+$, this case describes a regular black hole. 
From the standard geometries adopted here, it is worth pointing out that this case occurs only 
in the extremal Reissner--Nordström metric where $A>0$ for $r<r_+$.\footnote{Note that 
from a generic extremal black hole, this would also be possible.}  
Regular black holes are possible just for this latter case in the approach adopted here. 

\subsection{$C \neq C_0$}
For $C \neq C_0$, we do not avoid the curvature singularity at $r_0$.
 But two alternatives are still possible: the function $B(r)$ has either another positive zero at $r=r_{++}<r_0$ or at
$r=r_{++}> r_0$. With  $r_{++}< r_0$, the alternative is a naked singularity. But for 
$r_{++}> r_0$, the numerator of (\ref{delta}) leads to $C=-4r_{++}^2/A(r_{++})<0$ and
\begin{equation}
r^2 A(r_{++})- r_{++}^2A(r)<0,
\end{equation} 
for $r_0 <r< r_{++}$. Thus, $B<0$  as $A>0$ and the metric is not Lorentzian. 
But in this case, for $r>r_{++}$, we have $A>0$ and $B>0$. 
In this sense, the maximal extension of the metric is given by defining a new radial coordinate, 
the proper length $\ell (r)$ \cite{MT}:
\begin{equation}
\frac{d \ell (r)}{dr} = \frac{1}{\sqrt{B(r)}}.
\label{ell}
\end{equation} 
When an appropriate integration constant is fixed in Eq. (\ref{ell}), the interval $r_{++}<  r < \infty$ could be 
mapped into $0 < \ell < \infty$. The analytic extension in the $(t,\ell, \theta,\phi)$ coordinates 
is made complete continuing that interval to
$-\infty <\ell < \infty$. Then two asymptotic flat regions are connected by a throat at 
$\ell =0$ or $r_{\text{thr}}=r_{++}$, that is, 
for this case we have a wormhole with throat at $r_{++}$. Indeed, in order for $r_{++}$ to be 
the wormhole throat, $r(\ell)$ should have a minimum at $r_{\text{thr}}=r_{++}$ or $\ell =0$.

In summary, from the numerator of (\ref{delta}),  $4r_{++}^2+C A(r_{++})=4r_0^2+C_0A(r_0)=0$.
Therefore, we see that if $r_{++}>r_0$, 
$C<C_0$ and we have a wormhole. But if $r_{++} < r_0$, then $C>C_0$ and a naked singularity is present.

\section{Deviation from the Schwarzschild metric}
\label{Sec-IV}
In this section, in the revised Deser--Woodard nonlocal gravity, 
I am looking for analytic spacetime geometries that are deviations from 
the Schwarzschild metric. 
From the metric term of the Schwarzschild spacetime
\begin{equation}
A(r) = 1- \frac{2M}{r},
\label{A_Schw}
\end{equation}
where $M$ is the mass parameter, then Eqs. (\ref{B_delta})-(\ref{delta}) lead to
\begin{equation}
B(r) = \left(1- \frac{2M}{r} \right)^2 \left( \frac{4 r^3+ C \left(r - 2M \right) }{4r (r - 3M)^2}\right),  
\label{B_Schw}
\end{equation}
with a double real root at $r_+=2M$.
In this case, according to (\ref{C0}), the value of the integration constant that 
avoids the curvature singularity at $r_0=3M$ is
\begin{equation}
C_0 = -108 M^2.
\end{equation} 
As pointed out in Sec. \ref{Sec-III}, for a metric function like (\ref{B_Schw}), 
a singular black hole solution is available. 

\begin{figure}
\begin{centering}
\includegraphics[trim=1.55cm 0.2cm 0.6cm 0cm, clip=true,scale=0.56]{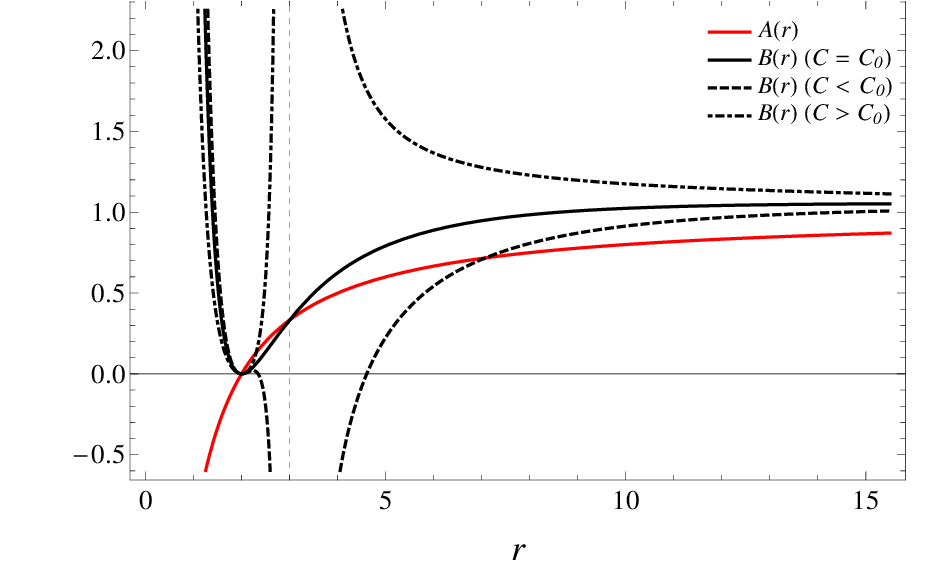}
\par\end{centering}
\caption{The metric functions given by Eqs.(\ref{A_Schw})-(\ref{B_Schw}) 
for the Schwarzschild deviation with different values of the integration
constant $C$. As one adopts $M=1$, the vertical dashed line indicates the curvature singularity at 
$r_0=3$, and the event horizon of the Schwarzschild black hole is located  at $r=2$.}
\label{BSchw}
\end{figure}

\subsection*{$C=C_0$ case}
 For $C=C_0$, the metric terms (\ref{A_Schw}) and (\ref{B_Schw})
describes a black hole. Spacetime is defined in the interval
\begin{equation}
0< r < \infty,
\label{interval1}
\end{equation}
and the function $B(r)$ is explicitly written as
\begin{equation}
B(r) = \left(1-\frac{2M}{r} \right)^2 \left(1+\frac{6M}{r} \right).
\label{singular_BH}
\end{equation}
 As we can see in Fig.  \ref{BSchw}, for $r<r_+$, 
the metric is not Lorentzian.
The metric becomes Lorentzian inside the horizon by using the quasiglobal coordinate $u$ 
and the definitions (\ref{QG_properties}). With the quasiglobal coordinate, the  
the radial coordinate $r(u)$ and metric term $\mathcal{A}(u)$ near the event horizon read
\begin{equation}
r(u) = 2M +\frac{u^2}{2M} + \mathcal{O}(u^4) \ \  \text{and} \ \ \mathcal{A}(u) \simeq 1- \frac{4M^2}{4M^2 + u^2}.
\end{equation}
At the event horizon $u=0$ ($r=2M$) and $\mathcal{A}(u)=0$. As $u<0$ inside the event horizon,
$\mathcal{A}(u)>0$, thus the metric is Lorentzian as expected. Also, this spacetime is singular, for the 
Kretschmann scalar
\begin{equation}
R_{\mu\nu\alpha\beta}R^{\mu\nu\alpha\beta} =  \frac{48 M^2}{r^6} -\frac{416 M^3}{r^7}+\frac{5520 M^4}{r^8}-\frac{19008 M^5}{r^9} +\frac{19008 M^6}{r^{10}}
\end{equation}
diverges just at $r=0$.

\subsection*{$C<C_0$ case}
For $C<C_0$, we have a wormhole with a wormhole throat localized at $r_{\text{thr}}=r_{++}$,
where $r_{++}$ is another real and positive root of the metric function $B(r)$ given by
\begin{equation}
r_{++} = -\frac{3 C - \left[C \left(54 M-3 \sqrt{3} \sqrt{C+108 M^2}\right)\right]^{\frac{2}{3}}}{6 \left[C \left(54 M-3 \sqrt{3} \sqrt{C+108 M^2}\right)\right]^{\frac{1}{3}}}.
\end{equation}
As we can see in Fig. \ref{BSchw}, $r_{++}>r_0$, and the radial coordinate is well defined in the interval
\begin{equation}
r_{++}< r < \infty.
\end{equation}
This  wormhole is traversable. By using the proper length (\ref{ell}), the analytic extension provides two asymptotic
flat regions connected by a wormhole throat at $r_{\text{thr}}= r_{++}$.

\subsection*{$C>C_0$ case}
For $C>C_0$,  the metric given by (\ref{A_Schw}) and (\ref{B_Schw}) leads to a naked singularity. Spacetime
metric is defined for $r>r_0=3M$. As the singularity is naked, there is no event horizon 
protecting the curvature singularity (see Fig. \ref{BSchw}).

\section{Deviation from the Reissner--Nordström metric}
\label{Sec-V}
Now, I am looking for analytic spacetime geometries that are deviations from 
the Reissner--Nordström metric. 
From the metric term of the Reissner--Nordström spacetime, namely
\begin{equation}
A(r) = 1- \frac{2M}{r} + \frac{Q^2}{r^2},
\label{A_RN}
\end{equation}
where again $M$ is the mass term, and $Q$ is the electric charge parameter, Eqs. (\ref{B_delta})-(\ref{delta}) lead to
\begin{equation}
B(r) =\bigg(1- \frac{2M}{r} + \frac{Q^2}{r^2} \bigg)^2 \bigg(\frac{4r^4 + C \left( r^2-2 Mr+Q^2\right)}{4 \left(r^2-3 Mr+2 Q^2\right)^2}\bigg),
\label{B_RN}
\end{equation}
with event horizon at $r_+ = M + \sqrt{M^2-Q^2}$ for $Q \leq M$. As expected, for $Q=0$ the metric functions are equal to the Schwarzschild functions given by 
Eqs.(\ref{A_Schw})-(\ref{B_Schw}). 

In the Reissner--Nordström deviation, the curvature singularity is at 
\begin{equation}
r_0 = \frac{1}{2} \left(3 M + \sqrt{9 M^2-8 Q^2}\right).
\label{r0_RN}
\end{equation}
And the value of $C$ that avoids such a singular point is
\begin{equation}
C_0 = \frac{\left(3 M + \sqrt{9 M^2-8 Q^2}\right)^4}{4 Q^2 - 2 M \left(3 M + \sqrt{9 M^2-8 Q^2}\right)}.
\end{equation}
All cases available in Sec. \ref{Sec-IV} are found in the Reissner--Nordström deviation. 
The only exception is a regular black hole for $Q=M$. In the general
relativity context, this condition  gives rise to the extremal Reissner--Nordström
black hole. The extremal Reissner--Nordström spacetime is an interesting
black hole because its Hawking temperature is zero. 
From the metric components (\ref{A_RN})-(\ref{B_RN}) 
and  the definition of the Hawking temperature \cite{Hawking:1975vcx}, in terms
of the surface gravity $\kappa$, one has 
\begin{equation}
T = \frac{\kappa}{2\pi} = \frac{1}{4 \pi} \sqrt{\frac{B(r)}{A(r)}} A'(r) \Bigg |_{r=r_+} = 0,
\end{equation}
even when $Q \neq M$. 
It is worth noting that the singular black hole of the previous section, given by Eqs. (\ref{A_Schw}) and (\ref{singular_BH}), 
 is also an extremal black hole, according to the Hawking temperature definition.  
 
\subsection*{$C=C_0$ case}
\begin{figure}
\begin{centering}
\includegraphics[trim=1.2cm 0cm 1cm 0.12cm, clip=true,scale=0.58]{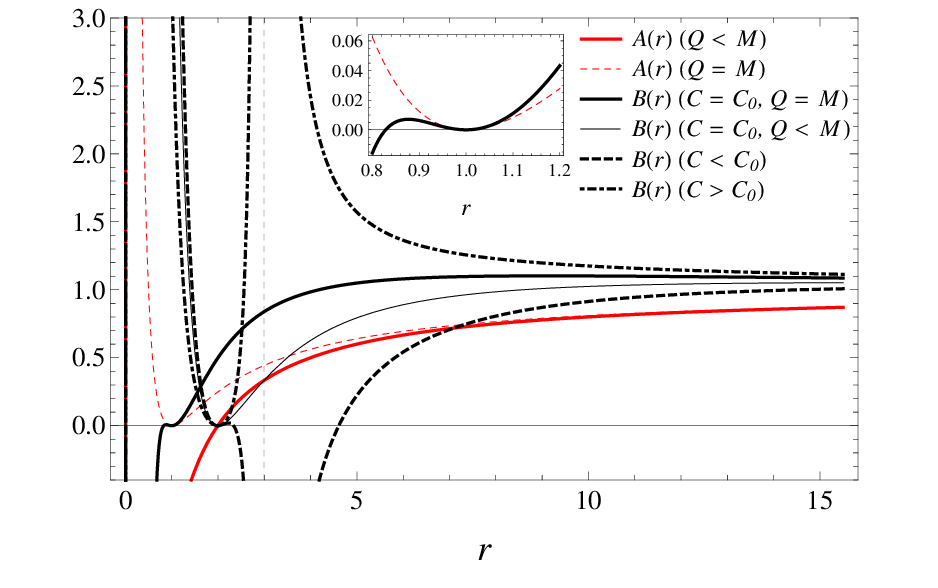}
\par\end{centering}
\caption{The metric functions given by Eqs.(\ref{A_RN})-(\ref{B_RN}) 
for the Reissner--Nordström deviation with different values of the integration
constant $C$. As one adopts $M=1$ and $Q=0.1$ (nonextremal), 
the vertical dashed line indicates the curvature singularity $r_0=2.99$ (nonextremal case). The event horizon
of the Reissner--Nordström black hole is located  at $r_+=1.99$ (nonextremal) and at $r_+=1$ (extremal). 
The smaller root of $B(r)$ is $r_- = 0.83$ for the extremal case (see the enlarged image).}
\label{BRN}
\end{figure}

 For $C=C_0$ and $Q<M$, the metric terms (\ref{A_RN}) and (\ref{B_RN})
also describes a singular black hole, whose spacetime is defined in the interval
\begin{equation}
0 < r < \infty.
\label{interval2}
\end{equation}
As we can see in Fig.  \ref{BRN}, for $r<r_{+}$, 
the metric is not Lorentzian, thus the valid interval of the radial coordinate is $r>r_+$.
In this sense, the analytic metric extension is made by using the the quasiglobal coordinate $u$, with the definitions
(\ref{QG_properties}), and the same successful result of the singular black hole, 
which is a deviation of the Schwarzschild metric, is
obtained for the Reissner--Nordström case.

In the extremal case, $Q=M$, consequently $r_+=M$, $C_0=-64M^2$, and the metric function $B(r)$ presents another positive 
root, given by
\begin{equation}
r_- =  2(\sqrt{2}-1)M, 
\end{equation}
that is, $r_-<r_+$. As mentioned before, $A(r)>0$ for $r_- < r < r_+$ like the function $B(r)$
 (see the enlarged graphic in Fig.  \ref{BRN}). Then, as we will see, spacetime is regular for
\begin{equation}
r_- \leq r < \infty.
\end{equation}
Therefore, as $r_+$ is the event horizon, one concludes 
that a regular black hole is possible for these values of $C$ and $Q$.
For the deformations studied in this article, this is the only case in which a regular black hole appears. As mentioned,
it is an extremal regular black hole with zero Hawking temperature.

\subsection*{$C<C_0$ case}
For $C<C_0$, a wormhole metric is obtained with throat localized at $r_{\text{thr}}=r_{++}$,
where again $r_{++}$ is another real and positive root of the metric function $B(r)$. Unfortunately, 
the numerator of $B(r)$ yields a fourth-order polynomial, whose expression is very large. For this
reason, I do not write such a root here. Most importantly, $r_{++}> r_0$,  the singularity
at $r=r_0$ is avoided. And, as we can see in Fig. \ref{BRN}, the spacetime metric is Lorentzian  for $r>r_{++}$, 
and the radial coordinate is valid for
\begin{equation}
r_{++}< r < \infty.
\end{equation}
The analytic extension is available with the new radial coordinate $\ell$. 
The interval above is mapped into the interval $0 < \ell < \infty$, and the maximal extension 
is obtained for $- \infty < \ell  < \infty$. Then again one has two asymptotic flat regions connected by the 
wormhole throat at $r_{++}$.  

\subsection*{$C>C_0$ case}
For $C>C_0$,  the metric given by (\ref{A_RN}) and (\ref{B_RN}) also leads to a naked singularity. Spacetime
geometry is defined for $r>r_0$, where $r_0$ is indicated in Eq. (\ref{r0_RN}). 
Again, there is no event horizon protecting the curvature singularity 
(see Fig. \ref{BRN}).

\section{Two interesting solutions}
\label{Sec-VI}
In this section, I will focus on two solutions and calculate the corresponding auxiliary fields
in order to check the correctness of the approach adopted here. Moreover, a brief discussion
on the energy conditions will be made.  

\subsection{Morris--Thorne wormhole}
The Morris--Thorne wormhole \cite{MT} is described by the metric 
\begin{equation}
ds^2= -e^{2\Phi(r)}dt^2+\left(1-\frac{b(r)}{r}\right)^{-1}dr^2+r^2 (d\theta^2 + \sin^2 \theta d\phi^2),
\label{Morris-Thorne}
\end{equation}
where $\Phi(r)$ is the redshift function, and $b(r)$ is the shape function.
In the simplest case, a case in which spacetime is horizonless and there are no tidal forces, the metric functions are 
$\Phi (r) =  0$ and $b(r)>0$. This specification is available in the approach developed in this article.
Making $M=0$ in Eq. (\ref{A_Schw}), consequently $C_0=0$, 
the deviation from Schwarzschild (\ref{B_Schw}) give us\footnote{This
solution for $\mathcal{C}=-1$ was obtained by D'Agostino and De Falco \cite{DAgostino:2025sta}.} 
\begin{equation}
ds^2 = - dt^2 + \left(1+ \frac{\mathcal{C}}{r^2} \right)^{-1}dr^2 +r^2 (d\theta^2 + \sin^2 \theta d\phi^2).
\label{MT}
\end{equation} 
Therefore, $\Phi (r) = 0$ and $b(r) = - \mathcal{C}/r$. In order to have a 
traversable wormhole with throat at $r_{\text{thr}}=\sqrt{- \mathcal{C}}$, the constant 
$\mathcal{C}$ should be negative. 

Another term in the field equations (\ref{field_equations}), 
namely $\mathcal{K}_{\mu\nu}$, 
could be calculated from the known metric terms and the function $W$. The nonzero
component of $\mathcal{K}_{\mu\nu}$ that appears in the field equations (for radial
 auxiliary fields) diverges at the wormhole throat and is written as
\begin{equation}
\mathcal{K}_{rr}= -\frac{2\mathcal{C}}{r^4+\mathcal{C}r^2}.
\label{k_MT}
\end{equation} 
However, as mentioned in Sec. \ref{Sec-II},
one should check this result using the auxiliary fields and the definition (\ref{k}). 

\begin{figure}
\begin{centering}
\includegraphics[trim=1.5cm 0.3cm 0.3cm 0cm, clip=true,scale=0.6]{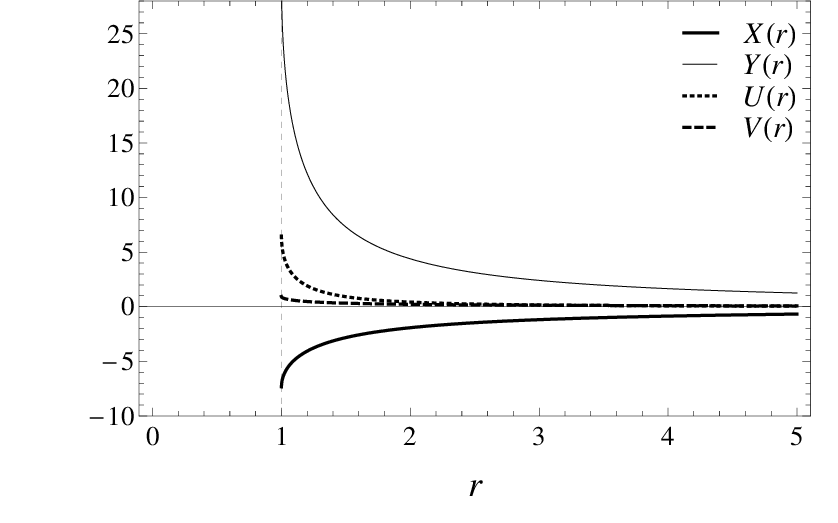}
\par\end{centering}
\caption{Auxiliary fields for the Morris--Thorne solution (\ref{MT}) 
with $\mathcal{C}=-1$. The vertical dashed line indicates the wormhole throat.
As we can see, all fields are finite at the throat. }
\label{MT_fields}
\end{figure}

With the metric (\ref{MT}), the Ricci scalar is $R=2 \mathcal{C}/r^4$, and
the auxiliary fields (\ref{X})-(\ref{V}) are straightforward:
\begin{subequations}
\begin{align}
X (r) = & - \frac{\ln ^2 \mathcal{C}}{4} + \left( \mathcal{F}(r) - \ln r  \right)^2, \label{XMT} \\
Y(r) = & \frac{1}{3} \big( \mathcal{G}(r)-\ln r \big) \big(2 \mathcal{F}(r) -\mathcal{G}(r) -\ln r \big) \big(2 \mathcal{F}(r)^2 - 2 \mathcal{F}(r) \mathcal{G}(r) - 2 \mathcal{F}(r) \ln r + \mathcal{G}(r)^2 + \ln ^2 r \big), \label{YMT} \\
U(r) = & \frac{6}{25} \bigg(\frac{32 (\mathcal{F}(r) - \ln r)^5}{\ln ^5  \mathcal{C}} -5 \ln [\mathcal{F}(r) - \ln r] 
 + 5 \ln \left[\frac{\ln  \mathcal{C}}{2}\right] -1\bigg), \label{UMT}  \\
V(r) = & - \frac{3}{10}  \left(\frac{32 (\mathcal{F}(r) - \ln r )^5}{\ln^5  \mathcal{C}}-1\right) \bigg(\mathcal{F}(r) - \ln r \bigg)^{-2} \label{VMT},
\end{align}
\end{subequations}
with 
\begin{equation}
\mathcal{F}(r) =  \ln \left[\sqrt{ \mathcal{C} ( \mathcal{C}+r^2)}+ \mathcal{C} \right] \ \ \ \text{and} \ \ \ \mathcal{G}(r) =  \ln \left[\sqrt{ \mathcal{C}+r^2}+\sqrt{ \mathcal{C}}\right].
\end{equation}
It is worth pointing out that the integration constants of the the auxiliary fields were set to either zero or the corresponding
values for the expected asymptotic behavior, that is, all auxiliary fields should be zero at infinity. For real-valued fields,
the integration constant $\mathcal{C}$ should be $-1$, like the case studied in Ref. \cite{DAgostino:2025sta}.
Interestingly, all auxiliary fields are finite at the wormhole throat (see Fig. \ref{MT_fields}), 
but, as mentioned, the term $\mathcal{K}_{rr}$ is 
not. $X$ is negative definite, and $Y>0$ like the cosmological context studied in Ref. \cite{Deser:2019lmm}.

Distant from the wormhole throat, the distortion function, given by Eq. (\ref{f}), reads
\begin{equation}
f(Y) \simeq  - \frac{432 Y^2}{\pi^8}.
\end{equation}
As $\lim_{r \rightarrow \infty}Y=0$, the distortion function also vanishes in this limit. 

From the auxiliary fields above, the definition of $\mathcal{K}_{\mu\nu}$, given by (\ref{k}), leads to the 
same result of (\ref{k_MT}). 
Therefore, the Morris--Thorne wormhole (\ref{MT}), with the auxiliary fields (\ref{XMT})-(\ref{VMT}) 
or $\mathcal{K}_{rr}$ given by  Eq. (\ref{k_MT}) and the condition $W=A^{-1}$, is solution 
of the vacuum field equations (\ref{field_equations}) in the  Deser--Woodard gravity. 

The last point to investigate regards energy conditions violations. In the Deser--Woodard context this is not the 
case because the Morris--Thorne wormhole is a vacuum spacetime.
But this is not the case in the general relativity context. Considering the field equations (\ref{field_equations}) 
as the Einstein equations, $G_{\mu\nu}=\kappa T_{\mu\nu}$, one defines
an effective energy-momentum tensor as
\begin{align}
T_{\mu\nu}= & \frac{1}{\kappa W}\big( \nabla_{\mu}\nabla_{\nu}W - g_{\mu\nu} \square W -\mathcal{K}_{(\mu\nu)} + \frac{1}{2}g_{\mu\nu}g^{\alpha \beta} \mathcal{K}_{\alpha \beta}\big),
\label{Eff}
\end{align}
where $\kappa=8\pi$. Assuming that this energy-momentum tensor reads 
\begin{equation}
T_{\nu}^{\mu}=\frac{1}{\kappa}\left(\begin{array}{cccc}
-\rho\\
 &p_r \\
 &  & p_t \\
 &  &  & p_t
\end{array}\right),
\label{Eff2}
\end{equation}
in which $\rho$ is the fluid energy density, $p_r$ is the radial pressure, and $p_t$ is the tangential
pressure, we have an anisotropic fluid, whose parameters are 
\begin{equation}
\rho=p_r= - p_t=\frac{\mathcal{C}}{\kappa r^4}. 
\end{equation} 
As the weak energy condition says that $\rho + p \geq 0$, for the effective energy-momentum tensor (\ref{Eff2}) one has
\begin{equation}
\rho + p_r = \frac{2\mathcal{C}}{\kappa r^4} < 0,
\end{equation}
because $\mathcal{C}<0$. Also, both the energy density and the radial pressure 
are negative. Interestingly, this is in agreement with the very purpose of the Deser--Woodard
model, which is to generate a dark energy phase without a cosmological constant. From the effective 
perspective, an exotic matter (like dark energy) supports a wormhole spacetime.

\subsection{Regular black hole} 
As I said, from the approach adopted here, the only fully analytic regular black hole solution in the Deser--Woodard 
gravity is the one described in Sec. \ref{Sec-V}. Its metric is explicitly written as
\begin{equation}
ds^2 =  -\bigg(1- \frac{M}{r} \bigg)^2dt^2 +\bigg[\bigg(1- \frac{M}{r} \bigg)^2  \bigg(1+\frac{4M}{r}-\frac{4M^2}{r^2} \bigg) \bigg]^{-1} dr^2 +r^2 (d\theta^2 + \sin^2 \theta d\phi^2).
\label{RBH}
\end{equation}
It is a regular black hole, because the event horizon is located at $r_+ = M$, and  as
the radial coordinate range is $r_- = 2(\sqrt{2}-1)M \leq r <\infty$, the singularity at $r=0$ is avoided.
According to the Kretschmann scalar for this spacetime, which is written as
\begin{equation}
R_{\mu\nu\alpha\beta}R^{\mu\nu\alpha\beta} = \frac{48 M^2}{r^6}-\frac{192 M^3}{r^7}+\frac{1656 M^4}{r^8}  -\frac{5936 M^5}{r^9} +\frac{9232 M^6}{r^{10}}-\frac{6528 M^7}{r^{11}}+ \frac{1728 M^8}{r^{12}}, 
\end{equation}
we see that the curvature singularity occurs only at $r=0$, which is not part of the spacetime metric (\ref{RBH}).

Indeed, the internal structure of this regular spacetime metric is of a wormhole with a throat at 
$r=r_-$. If we write the metric (\ref{RBH}) as a Morris--Thorne wormhole (\ref{Morris-Thorne}), 
the shape function $b(r)$ reads
\begin{equation}
b(r) = -2 M +\frac{11 M^2}{r}-\frac{12 M^3}{r^2} + \frac{4 M^4}{r^3},
\label{b(r)}
\end{equation}
and $b(r_-)=r_-$. Following Morris--Thorne \cite{MT}, 
the condition for a throat connecting two asymptotically flat 
spacetimes is given by the flaring-out condition:
\begin{equation}
\frac{b(r)-r b'(r)}{b(r)^2}>0
\label{Flaring-out}
\end{equation} 
at or near the throat. At the wormhole throat, this condition requires $b'(r_-)<1$.
That is the case for the shape function (\ref{b(r)}). In particular,
at the throat Eq. (\ref{Flaring-out}) equals $\frac{1}{2\sqrt{2}M}$ and $b'(r_-)=\frac{1}{\sqrt{2}}$. Therefore, one has
a regular black hole with a wormhole structure inside the event horizon or a wormhole throat
protected by an event horizon (a nontraversable wormhole).

\begin{figure}
\begin{centering}
\includegraphics[trim=0.6cm 0.3cm 0.3cm 0cm, clip=true,scale=0.6]{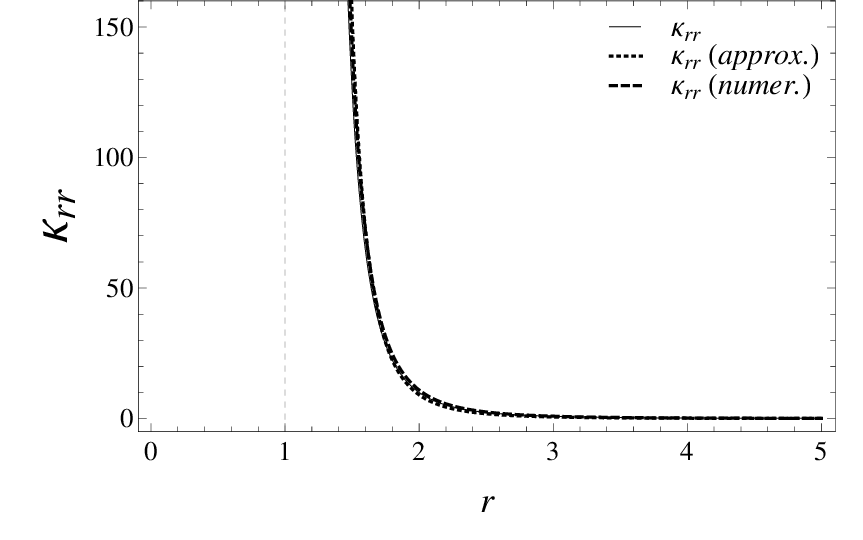}
\par\end{centering}
\caption{The tensor  component $\mathcal{K}_{rr}$, 
which depends on the auxiliary fields, calculated directly from
 the fields equations, approximately and numerically. 
 As one adopts $M=1$, the vertical dashed line indicates the event horizon. As
we can see, $\mathcal{K}_{rr}$ diverges at this point. For the approximate result, one adopts 
$X_0 =Y_0 = V_0 =2$. And for the numerical calculations, one assumes the asymptotic behavior of the auxiliary
fields. }
\label{kRBH}
\end{figure}

The term that involves the auxiliary fields in the field equations Eq. (\ref{field_equations}) is obtained from the 
known metric and the very assumption $W=A^{-1}$. This leads to
\begin{equation}
\mathcal{K}_{rr} = \frac{2 M^2 \left(19 r^2 - 20Mr + 4 M^2 \right)}{(r-M)^4 \left(r^2 + 4 M r - 4 M^2 \right)}.
\label{k_RBH}
\end{equation}
Here this component of the field equations diverges at the event horizon.
Unfortunately, analytic solutions for the auxiliary fields (\ref{X})-(\ref{V}) 
are not available for checking the result above. In order to circumvent this problem,
we should calculate the auxiliary fields approximately. From the Ricci scalar for the metric (\ref{RBH}),
given by $R=4 M^2 (r-M)(2 M-5 r)/r^6$, one assumes that $R = 0$ outside the horizon, 
leading to the homogeneous equation $\square X =0$ (for large values of $r$, this is a reasonable 
choice because $R \sim 1/r^4$). 
Thus, the auxiliary field
equations yield an approximate value for $\mathcal{K}_{rr}$, namely
\begin{equation}
\mathcal{K}_{rr} \simeq \frac{ V_0 r^2 \left[M Y_0 - X_0^2 (1 - \ln 16)\right]}{M (r-M)^4 \left(r^2 + 4 M r - 4 M^2\right)},
\end{equation}
where $X_0,Y_0$ and $V_0$ are integration constants of the auxiliary fields (other constants were set to zero
to obtain the expected asymptotic behavior at infinity). Here I omit the expressions for the approximate 
auxiliary fields due to large sizes. As we can see in Fig. \ref{kRBH}, for suitable choices for the integration
constants, the exact value of $\mathcal{K}_{rr}$ (extracted from the field equations) 
and its approximation (alongside the numerical calculation) are
in very good agreement.  In this sense, one concludes that the metric (\ref{RBH}) is solution of the vacuum
field equations (\ref{field_equations}).

Also, the distortion function $f(Y)$ could be evaluated approximately. For small values of the field $Y$, that is,
distant from the event horizon $(r\gg r_+ )$, the distortion function reads
\begin{equation}
f(Y) \simeq  - \frac{2M^2 Y}{M Y_0 - X_0^2 (1- \ln 16)},
\end{equation}
where $X_0$ and $Y_0$ are integration constants of the fields $X$ and $Y$, respectively. 
For large values of the radial coordinate, $Y \simeq 0$, thus $f(Y) \simeq 0$.

\begin{figure}
\begin{centering}
\includegraphics[trim=1.5cm 0.3cm 0.3cm 0.02cm, clip=true,scale=0.6]{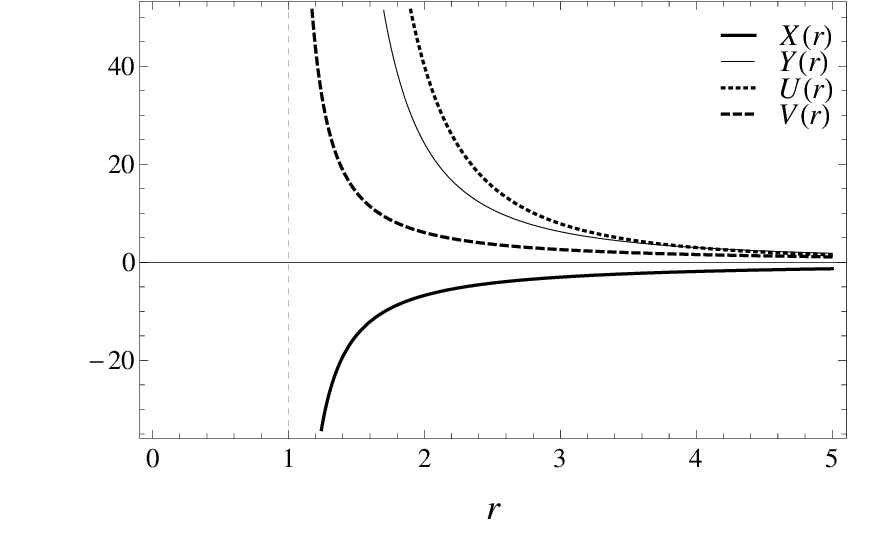}
\par\end{centering}
\caption{Auxiliary fields for the regular black hole solution (\ref{RBH}) numerically calculated. 
The vertical dashed line indicates the event horizon.
As we can see, all fields diverges at the event horizon. In this graphic, one adopts $M=1$. }
\label{RBH_fields}
\end{figure}

Each auxiliary field is numerically calculated in Fig. \ref{RBH_fields}. As we can see, the auxiliary fields diverge
at the event horizon or $r_+=M$. Also the approximate calculation indicates that the auxiliary 
fields are $\sim \frac{1}{r-M}$. The reason for that could be the constraint (\ref{W}), once that $A(r_+)=0$. 
But this is not the case. For example, the \enquote{Schwarzschild solution} in this theory, i.e., $A(r)=B(r)=1-2M/r$, 
gives us $W=1$, according to Eq. (\ref{G1}) or Eq. (\ref{G2}), with auxiliary fields, 
from Eqs(\ref{X})-(\ref{V}), written as 
\begin{subequations}
\begin{align}
X(r) = & \frac{X_0}{2M} \ln \left[ \frac{r-2M}{r}\right], \\
Y(r) = & \frac{X_0^2}{8M^2}\left( \ln \left[ \frac{r}{r-2M}\right]\right)^2, \\
U(r) = & - \frac{X_0 V_0}{4M^2}\left( \ln \left[ \frac{r}{r-2M}\right]\right)^2, \\
V(r) = & \frac{V_0}{2M} \ln \left[ \frac{r-2M}{r}\right],
\end{align} 
\end{subequations}
and distortion function given by
\begin{equation}
f(Y)=\frac{2V_0 Y}{X_0},
\end{equation}
where $X_0$ and $V_0$ are integration constants of the fields. 
Obviously, according to Eqs.(\ref{Gtt})-(\ref{Gtheta}) and $G_{\mu\nu}=0$, 
$\mathcal{K}_{\hat{r}\hat{r}}$ should be zero (and that is the case adopting the fields above). 
As we can see, all auxiliary fields diverge at the event horizon $r=2M$, 
then the existence of an event horizon would be cause of this divergence, not the constraint (\ref{W}).
The question is: are nonlocality and an event horizon incompatible? D'Agostino and 
De Falco \cite{DAgostino:2025wgl} argue that the auxiliary fields are finite at the event horizon,
but their spacetime metrics are approximations in the perturbation parameter. Such 
black hole metrics are not analytic as the regular black hole metric (\ref{RBH}). 

The divergence of the auxiliary fields for the regular black hole deserves further
investigation. This divergence may be due to a coordinate artifact or not. In order to remove this problem,
a different tetrad frame or even redefining the distortion function are alternatives.\footnote{These alternatives for a future work were suggested by an anonymous referee.}

As for the sign of the auxiliary fields, for the regular black hole, according to Fig. \ref{RBH_fields}, 
$X$ is negative definite, but $Y$ is positive definite. 
This is not in agreement with the argument given by Deser--Woodard \cite{Deser:2019lmm}, according to which,
for gravitationally bound systems $Y$ should be negative.

Lastly, let us examine the weak energy violation for the regular black hole. Assuming the general relativity context and 
an effective energy-momentum tensor like (\ref{Eff}) and (\ref{Eff2}), one has
\begin{subequations}
\begin{align}
\rho = & -\frac{M^2 \left(11 r^2 -24 M r+12 M^2\right)}{\kappa r^6}, \\
p_r = & \frac{M \left(4 r^3 -5M r^2 - 4 M^2 r + 4 M^3\right)}{\kappa  r^6}, \\
p_t = & -\frac{M \left(2 r^3 - 7 M r^2  + 4 M^3 \right)}{\kappa  r^6}.
\end{align} 
\end{subequations}
The energy density and the radial pressure show that
\begin{equation}
\rho + p_r = \frac{4M (r - M)^2 (r - 2 M)}{\kappa  r^6} < 0
\end{equation}
for $r<2M$. That is, the weak energy violation occurs even outside the event horizon, which is $r_+ = M$.
In particular, the energy density is negative for $r > 2(6+\sqrt{3})M/11$ (outside the event horizon) 
and positive otherwise.
Like the Morris--Thorne wormhole, from the general relativity context, there is an exotic matter field 
that gives rise to the spacetime metric. In particular, regular black holes are usually known for violating
energy conditions in general relativity \cite{Neves:2014aba}.  For known regular black hole
solutions like the Bardeen black hole, the source of the energy conditions violation could be either a nonlinear  
electrodynamics \cite{Ayon-Beato:2000mjt} or quantum aspects \cite{Maluf:2018ksj}.

\section{The shadow angular radius}
\label{Sec-VII}
\begin{figure}
\begin{centering}
\includegraphics[trim=0.4cm 0.3cm 0cm 0cm, clip=true,scale=0.55]{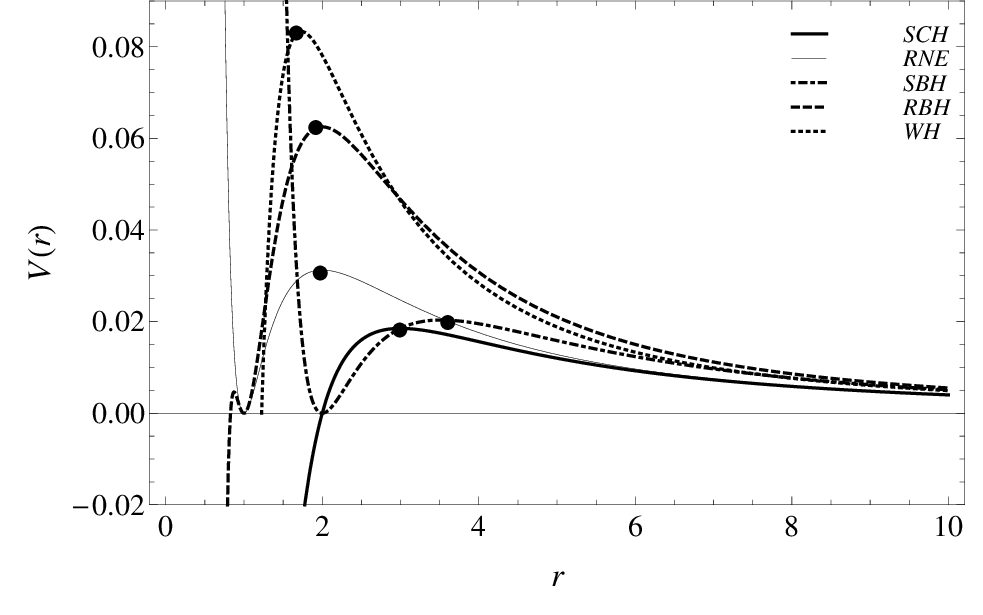}

\includegraphics[trim=0cm 0.3cm 1.7cm 0cm, clip=true,scale=0.56]{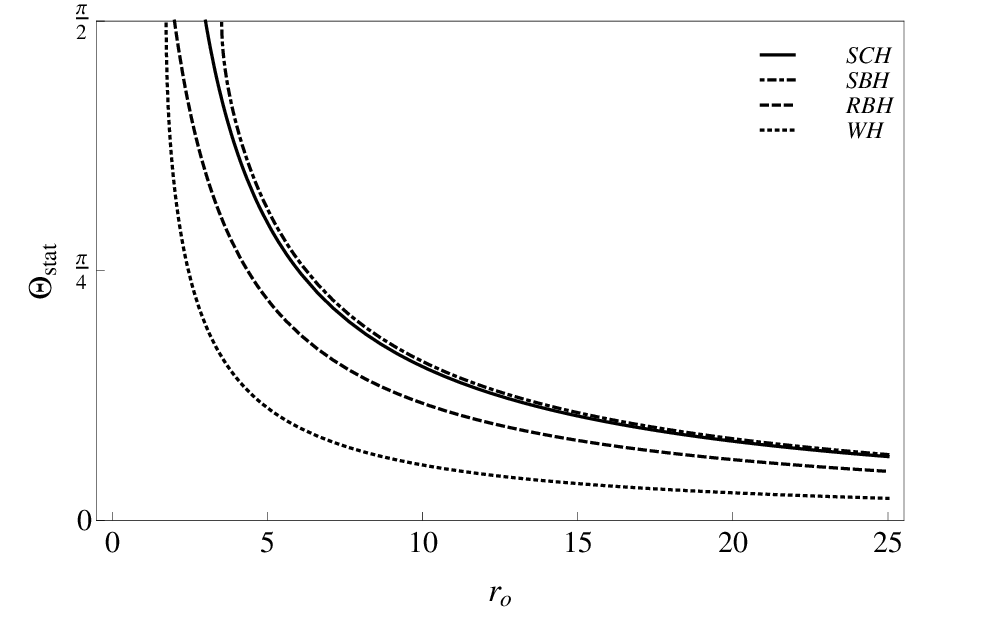}
\par\end{centering}
\caption{On the top: the gravitational potentials of the Schwarzschild black hole (SCH), 
extremal Reissner--Nordström (RNE), singular black hole (SBH), regular black hole (RBH) and wormhole (WH).
The last three come from the Deser--Woodard gravity. The dots indicate the photon sphere radius.
On the bottom: the shadow angular radius of the mentioned spacetime metrics (the regular black hole is equal to
the extremal Reissner--Nordström case) for a static observer at $r=r_o$. 
In this graph, one adopts $M=L=1$ and $\mathcal{C}=-1.5.$ }
\label{V-Theta}
\end{figure}

The shadow phenomenon, whether produced by a black hole or a wormhole, is a dark region in the sky caused 
by the gravitational light deflection. The shadow silhouette is given by the photon sphere, which is 
a region of unstable orbits for light rays.  
For the two interesting spacetime metrics of the last section and for the singular black hole
given by (\ref{A_Schw}) and (\ref{singular_BH}),\footnote{Like the regular case, the singular black
hole does not provide analytic expressions for all auxiliary fields. However, by using the same 
procedure of the regular case, one can see that the spacetime metric is solution of the field
equations in the Deser--Woodard context.}  the photon sphere radius is given by the following
conditions for the gravitational potential $V(r)$: $dV(r)/dr=0$ and $d^2 V(r)/dr^2<0$ 
 (calculated at the photon sphere radius $r=r_{ph}$). 
 According to Eq. (\ref{V-E}), the gravitational potential for the wormhole solution (WH), regular black hole (RBH),
and for the singular black hole (SBH) are
\begin{subequations}
\begin{align}
V_{WH}(r)= & L^2 \left(\frac{1}{2 r^2} + \frac{\mathcal{C}}{2 r^4}\right), \\
V_{RBH}(r) = & L^2 \left(\frac{1}{2 r^2} + \frac{M}{r^3}-\frac{11 M^2}{2 r^4}+\frac{6 M^3}{r^5} -\frac{2 M^4}{r^6}\right), \\
V_{SBH}(r) = & L^2 \left(\frac{1}{2 r^2}+\frac{M}{r^3}-\frac{10 M^2}{r^4}+ \frac{12 M^3}{r^5}\right),
\end{align} 
\end{subequations}
where $L$ is the photon angular momentum through null geodesics.
Thus the mentioned conditions for the potential lead to $r_{ph}=\sqrt{2\vert \mathcal{C} \vert}$ for the wormhole,
 $r_{ph}=2M$ for the regular black hole, and $r_{ph}=\frac{5}{2}(\sqrt{\frac{29}{5}}-1)M$ for 
the singular one. By comparison, the photon sphere
radius for the Schwarzschild black hole is $r_{ph}=3M$, and is $r_{ph}=2M$ 
for the extremal Reissner--Nordström black hole. As we can see, for a black hole with mass $M$, 
only the deviation from the Schwarzschild spacetime produces a different photon sphere (larger in this case)
when compared to the corresponding solution in general relativity. This will affect the shadow angular radius.
However, according to Fig. \ref{V-Theta}, the deviation from the extremal Reissner--Nordström
black hole, for the same value of $L$, yields a larger potential when compared to the corresponding general relativity
spacetime. Another point is that both black holes solutions in the Deser--Woodard gravity, 
as mentioned, are extremal ones, therefore there is a stable geodesic at the event horizon. For the the regular case,
there is an unstable orbit inside the event horizon, $r=(\sqrt{15}-3)M$, something absent in the corresponding
general relativity solution.

Following the general result obtained in Appendix \ref{Appendix}, namely Eq. (\ref{Theta}), 
for a static observer at $r=r_o$, 
the shadow angular radius for the wormhole, regular black hole and singular solutions in the 
Deser--Woodard context are written as
\begin{subequations}
\begin{align}
\sin^2 \Theta_{\text{stat}(WH)}= & \frac{ 2\vert \mathcal{C} \vert}{r_o^2},\\
\sin^2 \Theta_{\text{stat}(RBH)}= & \frac{16M^2}{r_o^2}\left(1-\frac{M}{r_o} \right)^2, \\
\sin^2 \Theta_{\text{stat}(SBH)}= & \frac{27M^2}{\left(\frac{7}{10}+\frac{1}{10}\sqrt{\frac{29}{5}} \right)r_o^2} \left(1-\frac{2M}{r_o}\right).
\end{align}  
\end{subequations} 
For distant observers ($r_0 \gg 3M$), the small angle approximation, $\sin \Theta_{\text{stat}} \approx \Theta_{\text{stat}}$, is valid. Then the shadow angular radius for the Schwarzschild black hole
is the known result: $\Theta_{\text{stat}(SCH)}=3\sqrt{3}M/r_o$. As for the wormhole angular radius,
for $\mathcal{\vert C \vert}=\mathcal{C}_{0}=27M^2/2$ we have the same angular radius of Schwarzschild's. With
$\mathcal{\vert C \vert}>\mathcal{C}_{0}$, we have a larger shadow, 
for $\mathcal{\vert C \vert} <\mathcal{C}_{0}$, a smaller one.  
For the regular black hole, 
within this approximation, $\Theta_{\text{stat}(RBH)}=4M/r_o$, which equals the shadow angular
radius of the extremal Reissner--Nordström black hole and is smaller than the
Schwarzschild angular radius. On the other hand, for the singular black hole, 
 $\Theta_{\text{stat}(SBH)}> \Theta_{\text{stat}(SCH)}$. And this is more evident 
 close to the photon sphere (see Fig. \ref{V-Theta}). Therefore, from the observational point of view, 
 a static singular black hole in the Deser--Woodard 
 gravity would have a larger shadow than the Schwarzschild black hole in the general relativity context.  
 In particular, for Sgr A*, a compact object in the center of the  Milky Way galaxy 
 with mass $M \approx 4\times 10^6 M_{\odot}$ at 
 $r_o = 8$ kpc from Earth, the EHT collaboration \cite{EventHorizonTelescope:2022wkp} 
 claims $d_{\text{sh}}=48.7\pm 7.0$ $\mu$as  for the observed shadow
 angular diameter.  Using the Schwarzschild metric and both the regular black hole and the singular black hole 
 in the Deser--Woodard theory, we have $d_{\text{SCH}}\approx 50$ $\mu$as,  
 $d_{\text{RBH}}\approx 39$ $\mu$as, and $d_{\text{SBH}}\approx 52$ $\mu$as, respectively. 
 As we can see, the singular black hole fits the data better in this approximation.

\section{Final remarks}
\label{Sec-VIII}
Following a recent approach \cite{DAgostino:2025sta,DAgostino:2025wgl}, I adopt a special tetrad frame to 
obtain fully analytic black hole and wormhole solutions in the revised Deser--Woodard theory of gravity.
From deviations of either the Schwarzschild metric or the Reissner--Nordström metric, this article shows
that regular black holes are viable only from the extremal Reissner--Nordström geometry. On the other hand,
singular black hole and wormhole solutions come from deviations of both standard spacetime geometries.

The shadow angular radius was obtained for black holes and wormholes. 
As for wormholes, their shadows could be larger or smaller than the Schwarzschild shadow.
On the other hand,  the regular black hole, which is a deviation from the Reissner--Nordström metric, 
casts the same shadow when it is compared with the standard spacetime in general relativity. 
But the deviation from the Schwarzschild black hole obtained casts a larger shadow than the 
corresponding black hole in general relativity. This represents an interesting difference from the observational
point of view.

As for the wormhole geometries, the auxiliary fields, which carry nonlocal effects, are finite throughout  
spacetime. Interestingly, regarding the regular black hole solution, spacetime is regular everywhere, but
the auxiliary fields diverge at the event horizon. This could be a serious caveat for the Deser--Woodard theory or,
at least, for the coexistence of nonlocality and horizons. 

\section*{Data availability statement}
This manuscript has no associated data.

\section*{Acknowledgments}
I thank Fernando Gardim and Rodrigo Cuzinatto for useful comments and suggestions. Also I thank the ICT-Unifal
for the kind hospitality. 
This work was supported by Conselho Nacional de Desenvolvimento Científico e Tecnológico (CNPq), Brazil, 
(Grant No. 170579/2023-9).

\section*{ORCID iDs}
Juliano C. S. Neves 0000-0003-0049-0209

\appendix
\section{The shadow phenomenon for general static spacetimes}
\label{Appendix}
The shadow phenomenon is caused by the photon sphere, whose silhouette is given by unstable photon orbits.
A small perturbation moves photons on the photon sphere away to the black hole event horizon or to a distant
observer. In order to calculate such orbits or geodesics, let us write down the Lagrangian for particles on
 geodesics:
 \begin{equation}
 \mathcal{L}= \frac{1}{2}g_{\mu\nu}\dot{x}^{\mu}\dot{x}^{\nu},
 \label{Lagrangian}
  \end{equation}
where dot means derivative with respect to the affine parameter $\tau$. 
For a general metric \textit{Ansatz} like (\ref{ansatz}), equatorial null geodesics $(\theta =\frac{\pi}{2})$ are 
solutions of (\ref{Lagrangian}) as $\mathcal{L}=0$, that is, 
\begin{equation}
-A(r) \dot{t}^2 + \frac{\dot{r}^2}{B(r)} + r^2 \dot{\phi}^2 = 0.
\label{Geodesics}
\end{equation}
As the general spherical \textit{Ansatz} presents the two Killing vector fields responsible for the energy 
conservation and the momentum conservation, $\partial / \partial t$ and $\partial / \partial \phi$, respectively,
the energy $E$ and the angular momentum $L$ of photons on these equatorial geodesics read
\begin{equation}
E=A(r) \dot{t} \ \ \  \mbox{and} \ \ \ L=r^2 \dot{\phi}.
\end{equation} 
From such relations, as it is known from textbooks, Eq. (\ref{Geodesics}) may be written as a familiar equation that describes a unitary mass
subject to a potential $V(r)$, i.e.,
\begin{equation}
\frac{1}{2}\left(\frac{dr}{d\tau}\right)^2 +V(r) = \mathcal{E}(r).
\end{equation}
Using (\ref{Geodesics}) and the definitions of $E$ and $L$, the terms $V(r)$ and $\mathcal{E}$ 
for a general static metric are
\begin{equation}
V(r) = \frac{L^2B(r)}{2r^2} \ \ \  \mbox{and} \ \ \ \mathcal{E}(r)= \frac{E^2B(r)}{2A(r)}.
\label{V-E}
\end{equation} 
The conditions $dV(r)/dr =0$ and $d^2 V(r)/dr^2<0$ yield the photon sphere radius $r_{ph}$.

Circular geodesics are calculated from $dr/d\phi=0$. Using this result in Eq. (\ref{Geodesics}), one has
\begin{equation}
\frac{E^2}{L^2}=\frac{A(r)}{r^2}.
\label{E-L}
\end{equation}
In particular, on the photon sphere $r=r_{ph}$, thus $E^2/L^2$ is constant.

\begin{figure}
\begin{centering}
\includegraphics[trim=0cm 0.3cm 0.3cm 0.02cm, clip=true,scale=0.6]{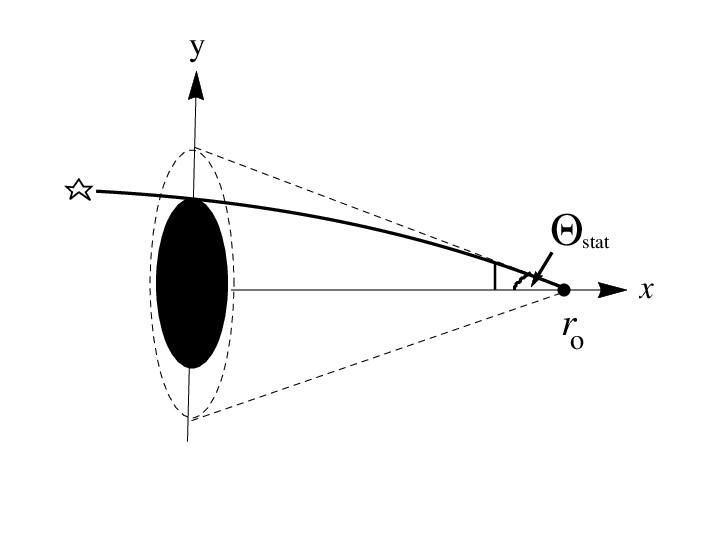}
\par\end{centering}
\caption{Representation of the shadow angular radius $\Theta_{\text{stat}}$ for a static observer
at $r=r_o$.}
\label{Representation}
\end{figure}

Following Ref. \cite{Maluf:2020kgf}, the shadow angular radius, 
according to Fig. \ref{Representation},  is measured by an observer 
outside the photon sphere, at $r=r_o>r_{ph}$, and is defined as
\begin{equation}
\tan \Theta = \lim_{\Delta x \rightarrow 0} \frac{\Delta y}{\Delta x}.
\end{equation}
The angular radius $\Theta$ is obtained by using the isotropic coordinates in such a way that 
angles are invariant as we compare them with angles in the Euclidean space. Thus the two-dimensional
spatial metric from the general static \textit{Ansatz} (adopting $\theta= \pi/2$) is conformal to the two-dimensional
Euclidean space with spherical coordinates $(\tilde{r},\phi)$, that is,  
\begin{equation}
ds^2_2 = \Omega (\tilde{r})^2 \left(d \tilde{r}^2 +\tilde{r}^2 d\phi^2 \right),
\end{equation}
with
\begin{equation}
r^2=\Omega(\tilde{r})^2 \tilde{r}^2 \ \ \ \mbox{and} \ \ \ \frac{dr^2}{B(r)}= \Omega(\tilde{r})^2 d\tilde{r}^2. 
\end{equation}
Here $\Omega(\tilde{r})$ is the conformal factor. From such relations and the definition of the shadow angular radius,
one has
\begin{equation}
\left(\frac{dy}{dx} \right)^2 = B(r)r^2 \left(\frac{d\phi}{dr} \right)^2.
\end{equation}
From the geodesics equation (\ref{Geodesics}), the term $d\phi/ dr$ is obtained. And with a useful trigonometric relation
$\sin^2 \Theta= \tan^2 \Theta/(1+\tan^2 \Theta)$, the desired result is achieved:
\begin{equation}
\sin^2 \Theta= A(r) \left(\frac{E}{L}r \right)^{-2}.
\end{equation}
For our static observer at $r=r_o$, the constant term $E^2/L^2$ comes 
from photons that orbited the photon sphere and is given by (\ref{E-L}) with $r=r_{ph}$. Therefore, for the 
general spherical symmetric spacetime, the shadow angular 
radius measured by our static observer is
\begin{equation}
\sin^2 \Theta_{\text{stat}}= \frac{A(r_o)}{A(r_{ph})}\frac{r_{ph}^2}{r_o^2}.
\label{Theta}
\end{equation}


\end{document}